\documentclass[aps,pre,
notitlepage
]{revtex4-1}

\usepackage{amsmath,amssymb,bm}
\usepackage{graphicx}
\usepackage{hyperref}

\begin{document}

\title{Counterexamples to the conjectured ordering between the waiting-time bound and the thermodynamic uncertainty bound on entropy production}

\author{Jie Gu}
\affiliation{Chengdu Academy of Educational Sciences, Chengdu 610036, China}

\date{\today}

\begin{abstract}
Two widely used model-free lower bounds on the steady-state entropy production rate of a continuous-time Markov jump process are the thermodynamic uncertainty relation (TUR) bound $\sigma_\text{TUR}$, derived from the mean and variance of a current, and the waiting-time distribution (WTD) bound $\sigma_\mathcal{L}$, derived from the irreversibility of partially observed transition sequences together with their waiting times.
It has been conjectured that $\sigma_{\mathcal L}$ is always at least as tight as  $\sigma_{\mathrm{TUR}}$ when both are constructed from the same partially observed link.
Here we provide four-state counterexamples in a nonequilibrium steady state where $\sigma_{\mathcal L}<\sigma_{\mathrm{TUR}}$.
This result shows that no universal ordering exists between these two inference bounds under partial observation.
\end{abstract}

\maketitle

\section{Introduction}

Entropy production quantifies nonequilibrium behavior and provides a universal measure of thermodynamic cost in small fluctuating systems. In stochastic thermodynamics, the steady-state entropy production rate $\sigma$ can be expressed exactly if the full Markov network and all transition rates are known. In many experimental situations, however, only coarse-grained observables are accessible, motivating model-free lower bounds on $\sigma$ \cite{seifert2025}.

Two prominent classes of such bounds are as follows. The thermodynamic uncertainty relation (TUR) yields a lower bound on $\sigma$ from the mean and variance of any time-integrated current in a nonequilibrium steady state \cite{barato2015,gingrich2016}. Separately, transition-based coarse-graining schemes that retain only a subset of observable transitions, together with their waiting times, yield a lower bound on $\sigma$ based on the Kullback--Leibler (KL) divergence between coarse-grained trajectories and their time reversals. This waiting-time distribution (WTD) bound has been developed in several complementary frameworks \cite{harunari2022,vandermeer2022}.

Because the WTD bound uses additional information beyond current statistics alone, it is natural to expect that it should be tighter than the TUR bound when both are inferred from the same partially observed link, as conjectured in Refs. \cite{harunari2022,vandermeer2022,seifert2025}. Indeed, for unicyclic networks, transition-based inference from a single observed link can recover the full entropy production, in which case $\sigma_{\mathcal L}=\sigma\ge \sigma_{\mathrm{TUR}}$ holds \cite{barato2015,gingrich2016}. For multicyclic networks, however, it has remained unclear whether a universal ordering $\sigma_{\mathcal L}\ge \sigma_{\mathrm{TUR}}$ holds.

In this work we resolve this question negatively by constructing an explicit continuous-time Markov jump process for which $\sigma_{\mathcal L}<\sigma_{\mathrm{TUR}}$.

\section{Setup and Notations}

\subsection{Continuous-time Markov jump process}

We consider a continuous-time Markov jump process on $N$ discrete states with transition rate matrix $W$ written in the column convention: $W_{ij}$ is the transition rate $j\to i$ for $i\neq j$, and each column sums to zero so that $W_{jj}=-\sum_{i\neq j}W_{ij}$. The stationary distribution $\bm p$ satisfies
\begin{equation}
W \bm p = \bm 0, \qquad \bm 1^{\mathsf T}\bm p = 1,
\end{equation}
with $\bm 1$ the all-ones vector.

The steady-state entropy production rate can be written as
\begin{equation}
\sigma = \sum_{i\neq j} p_j W_{ij}\,\ln\!\left(\frac{p_j W_{ij}}{p_i W_{ji}}\right),
\label{eq:sigma_full}
\end{equation}
which is nonnegative and vanishes if and only if detailed balance holds.

\subsection{Observed link and current}
For $N=3$, any connected network is unicyclic (a ring), and it is known that $\sigma_\mathcal{L} = \sigma_{\ell} \ge \sigma_\text{TUR}$.
Thus, we search for the minimal counterexamples in $4$-state networks.
We assume that only a single bidirectional transition pair between states $1$ and $4$ is observed. We denote the observed transition $4\to 1$ as ``$+$'' and $1\to 4$ as ``$-$''. From the observed time series we can construct a time-integrated current $X_T$ over a long observation time $T$ by counting $4\to 1$ as $+1$ and $1\to 4$ as $-1$, while all other transitions are unobserved and carry weight $0$.

In the long-time limit, the current has mean $j=\langle X_T\rangle/T$ and variance $\mathrm{Var}(X_T)\simeq 2DT$, defining the diffusion coefficient $D$.

%\section{Two lower bounds from the observed link}

%\subsection{TUR bound}

For continuous-time Markov jump processes in a nonequilibrium steady state, the TUR implies
\begin{equation}
\sigma \ge \sigma_{\mathrm{TUR}} \equiv \frac{j^2}{D}.
\label{eq:sigma_tur}
\end{equation}
This holds for any choice of current, including the partially observed current defined above.

%\subsection{Waiting-time bound from transition-based coarse-graining}

Next, we summarize the waiting-time  bound inferred from observing only the $+$ and $-$ transitions and the waiting times between consecutive observed events, following Refs. \cite{harunari2022,vandermeer2022}.

Let the steady-state observed transition rate (traffic on the observed link) be
\begin{equation}
h_K = p_4 W_{1,4} + p_1 W_{4,1}.
\end{equation}
The probability that an observed event is of type $+$ or $-$ is
\begin{equation}
P(+) = \frac{p_4 W_{1,4}}{h_K}, \qquad P(-)=\frac{p_1 W_{4,1}}{h_K}.
\end{equation}

Define the survival generator $S$ as the generator with the observed transition rates removed,
\begin{equation}
S = W - W_{1,4}\,|1\rangle\langle 4| - W_{4,1}\,|4\rangle\langle 1| ,
\label{eq:S_def}
\end{equation}
again in the column convention. Intuitively, $S$ governs the evolution between observed transitions, conditioned on no observed transition occurring.

The conditional probabilities that the next observed transition is the same direction as the previous one are given by
\begin{equation}
P(+|+) = -W_{1,4}\,(S^{-1})_{4,1}, \qquad
P(-|-) = -W_{4,1}\,(S^{-1})_{1,4},
\label{eq:cond_probs}
\end{equation}
where $(S^{-1})_{a,b}$ denotes the $(a,b)$ entry of $S^{-1}$, with indices referring to states.

The normalized waiting-time densities for successive same-direction observed transitions are
\begin{equation}
\psi_{++}(t) = -\frac{[e^{tS}]_{4,1}}{(S^{-1})_{4,1}}, \qquad
\psi_{--}(t) = -\frac{[e^{tS}]_{1,4}}{(S^{-1})_{1,4}}.
\label{eq:psi_pp_mm}
\end{equation}

The WTD bound can be written as a sum of an ``embedded-chain'' contribution and a waiting-time KL contribution,
\begin{equation}
\sigma_{\mathcal L} = \sigma_\ell + \sigma_t,
\end{equation}
with
\begin{equation}
\sigma_\ell
= h_K\,[P(+)-P(-)]\,\ln\!\left(\frac{P(+|+)}{P(-|-)}\right),
\label{eq:sigma_ell}
\end{equation}
and
\begin{equation}
\sigma_t
= h_K \Big[
P(+)P(+|+)\,D_{\mathrm{KL}}\!\left(\psi_{++}\Vert\psi_{--}\right)
+ P(-)P(-|-)\,D_{\mathrm{KL}}\!\left(\psi_{--}\Vert\psi_{++}\right)
\Big],
\label{eq:sigma_t}
\end{equation}
where $D_{\mathrm{KL}}(f\Vert g)=\int_0^\infty dt\, f(t)\ln[f(t)/g(t)]$.

For unicyclic networks, $\sigma_{\mathcal L}$ obtained from a single observed link can coincide with the full entropy production rate. For multicyclic networks, $\sigma_{\mathcal L}$ remains a valid lower bound on $\sigma$ but generally becomes strict.

\section{Counterexamples: $\sigma_{\mathcal L}<\sigma_{\mathrm{TUR}}$}

We generated $20000$ random transition rate matrices by drawing independent positive off-diagonal rates from a lognormal distribution, and calculated relevant quantities as described in Appendix \ref{app:detail}.
As seen from Fig. \ref{fig:search}, even though very rarely, instances of $\sigma_\mathcal{L} < \sigma_\text{TUR}$ exist ($2$ out of $20000$).
Instances of $\sigma_\ell < \sigma_\text{TUR}$ are more common ($924$ out of $20000$).

\begin{figure}[htb!]
\includegraphics[width=\columnwidth]{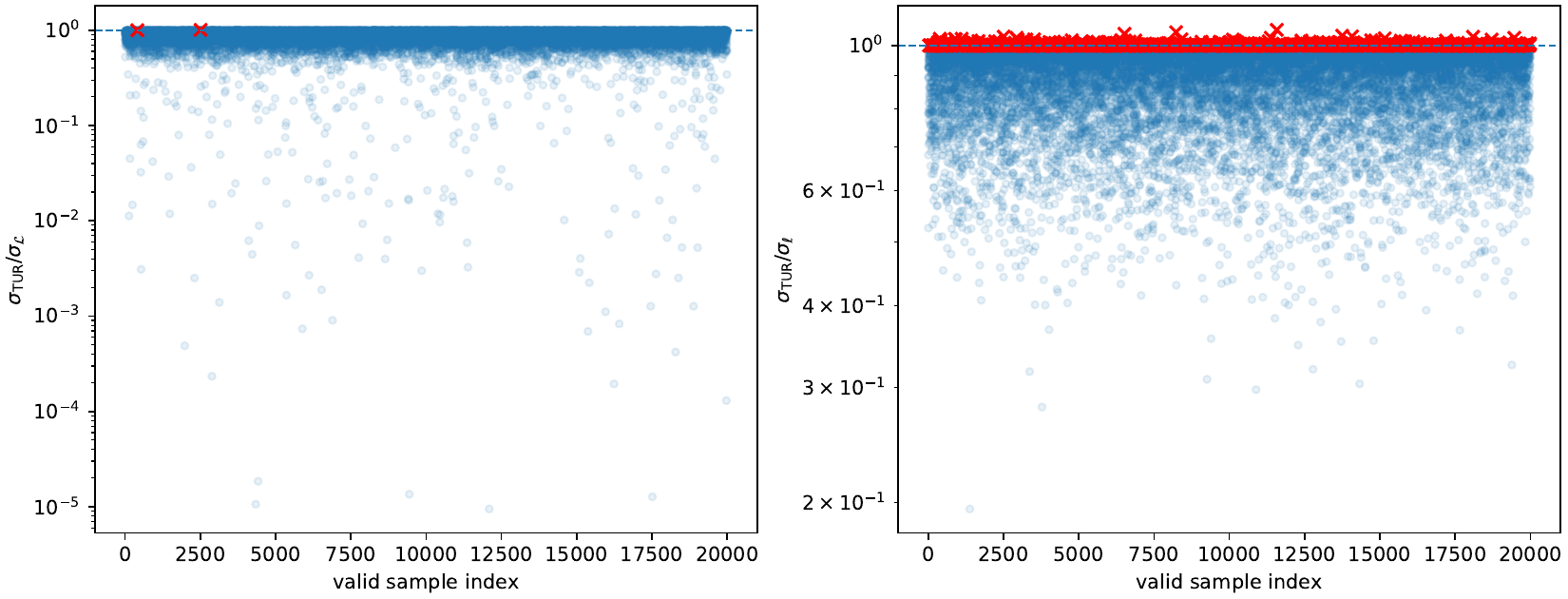}
\caption{Numerical search over randomly generated $4$-state Markov jump processes (all-to-all positive rates drawn from a lognormal distribution) with a single observed link $1\leftrightarrow 4$. The scatter plot shows (a) $\sigma_{\mathrm{TUR}}/\sigma_{\mathcal L}$ and (b) $\sigma_{\mathrm{TUR}}/\sigma_{\ell}$ versus the index of valid sampled generators. Points above the horizontal line at $1$ (represented by red crosses) indicate counterexamples with $\sigma_{\mathcal L}<\sigma_{\mathrm{TUR}}$ or $\sigma_{\mathcal L}<\sigma_{\ell}$.}
\label{fig:search}
\end{figure}

We also provide an explicit $4$-state transition rate matrix $W$ with particularly clean parameters:
\begin{equation}
W=\begin{pmatrix}
-58 & 3 & 37 & 24\\
11 & -50 & 3 & 51\\
5 & 27 & -44 & 4\\
42 & 20 & 4 & -79
\end{pmatrix}.
\label{eq:W_counterexample}
\end{equation}
All off-diagonal elements are positive and each column sums to zero, so Eq.~\eqref{eq:W_counterexample} defines a valid continuous-time Markov jump process.

The stationary distribution is
\begin{equation}
\bm p \approx (0.25355774,\;0.29519462,\;0.23006321,\;0.22118443)^{\mathsf T}.
\end{equation}
For the observed current counting $4\to 1$ as $+1$ and $1\to 4$ as $-1$, we obtain
\begin{equation}
j \approx -5.3410, \qquad D \approx 2.3093,
\end{equation}
leading to
\begin{equation}
\sigma_{\mathrm{TUR}}=\frac{j^2}{D} \approx 12.353.
\end{equation}
Evaluating Eqs.~\eqref{eq:sigma_ell} and \eqref{eq:sigma_t} for the same observed link yields
\begin{equation}
\sigma_\ell \approx 11.948,\qquad
\sigma_t \approx 0.256,\qquad
\sigma_{\mathcal L}\approx 12.204.
\end{equation}
Therefore,
\begin{equation}
\sigma_{\mathcal L} \approx 12.204 \;<\; 12.353 \approx \sigma_{\mathrm{TUR}},
\end{equation}
which directly refutes the conjectured universal ordering $\sigma_{\mathcal L}\ge \sigma_{\mathrm{TUR}}$.
We double-checked each intermediate result (see Appendix \ref{app:doublecheck}), confirming that this ordering is not due to numerical errors.. 

For completeness, the full entropy production rate computed from Eq.~\eqref{eq:sigma_full} is
\begin{equation}
\sigma \approx 40.99083759,
\end{equation}
so both $\sigma_{\mathcal L}$ and $\sigma_{\mathrm{TUR}}$ remain strict lower bounds on $\sigma$ in this example.

\section{Conclusion}

We provided an explicit continuous-time four-state Markov jump process for which the waiting-time bound $\sigma_{\mathcal L}$ inferred from a single observed bidirectional link is strictly smaller than the TUR bound $\sigma_{\mathrm{TUR}}$ inferred from the same link.
This counterexample shows that the conjectured universal ordering $\sigma_{\mathcal L}\ge \sigma_{\mathrm{TUR}}$ is false in general. It remains of interest to characterize subclasses of networks or observation schemes for which an ordering can be established.

\appendix

\section{Details of numerics}
\label{app:detail}

\subsection{Random sampling of the transition rate matrices}
We consider a continuous-time Markov jump process on $N=4$ states with transition rate matrix $W$ written in the column convention: for $i\neq j$, the off-diagonal element $W_{ij}>0$ is the transition rate for $j\to i$, and each column sums to zero,
\begin{equation}
\sum_{i=1}^N W_{ij}=0 \quad \text{for each } j.
\end{equation}
In numerical sampling, we draw independent positive off-diagonal rates from a lognormal distribution,
\begin{equation}
W_{ij}=\texttt{rate\_scale}\times \exp\!\big(\mu+\sigma\,\xi_{ij}\big), \quad \xi_{ij}\sim \mathcal{N}(0,1), \quad i\neq j,
\end{equation}
and then set the diagonal entries to enforce column conservation,
\begin{equation}
W_{jj}=-\sum_{i\neq j}W_{ij}.
\end{equation}
We have used $\texttt{rate\_scale}=1.0, \mu=0.0, \sigma =1.0$.

\subsection{Stationary distribution}
The stationary distribution $\bm p$ satisfies
\begin{equation}
W\bm p=\bm 0, \qquad \bm 1^{\mathsf T}\bm p=1,
\end{equation}
with $\bm 1$ the all-ones vector. Numerically, we solve the augmented linear system in a least-squares sense,
\begin{equation}
\begin{pmatrix}
W\\
\bm 1^{\mathsf T}
\end{pmatrix}
\bm p \approx
\begin{pmatrix}
\bm 0\\
1
\end{pmatrix},
\end{equation}
using a standard linear least-squares routine.
The tolerance has been set to $10^{-14}$.
 The resulting vector is projected onto the nonnegative orthant by setting negative components to zero and renormalized to sum to one. Samples for which the normalization fails (sum too small or non-finite) are rejected.

\subsection{Observed link current and its long-time variance}
We assume that only the bidirectional link $1\leftrightarrow 4$ is observed, and define a signed jump current that counts the transition $4\to 1$ as $+1$ and $1\to 4$ as $-1$. In column convention, $4\to 1$ corresponds to the matrix entry $(i,j)=(1,4)$, and $1\to 4$ corresponds to $(i,j)=(4,1)$. Using zero-based indices as in the implementation, these are $\texttt{PLUS}=(0,3)$ and $\texttt{MINUS}=(3,0)$.

The mean current $j$ and diffusion coefficient $D$ are computed via perturbation theory for the dominant eigenvalue of the tilted generator, avoiding numerical differentiation. Define the derivative matrices $W_1$ and $W_2$ by
\begin{equation}
(W_1)_{ij}=
\begin{cases}
W_{ij}, & (i,j)=(1,4),\\
-\,W_{ij}, & (i,j)=(4,1),\\
0, & \text{otherwise},
\end{cases}
\qquad
(W_2)_{ij}=
\begin{cases}
W_{ij}, & (i,j)=(1,4)\ \text{or}\ (4,1),\\
0, & \text{otherwise},
\end{cases}
\end{equation}
with all diagonal entries of $W_1$ and $W_2$ set to zero. The mean current is then
\begin{equation}
j=\bm 1^{\mathsf T}W_1\bm p.
\end{equation}
To obtain the second derivative of the scaled cumulant generating function at zero tilt, we solve a Poisson-type equation for an auxiliary vector $\bm r$,
\begin{equation}
W\bm r=-(W_1\bm p-j\bm p), \qquad \bm 1^{\mathsf T}\bm r=0,
\end{equation}
again using a least-squares solve of an augmented system with the constraint $\bm 1^{\mathsf T}\bm r=0$. The second derivative is evaluated as
\begin{equation}
\lambda''(0)=\bm 1^{\mathsf T}W_2\bm p + 2\,\bm 1^{\mathsf T}W_1\bm r,
\end{equation}
and the diffusion coefficient is
\begin{equation}
D=\frac{1}{2}\lambda''(0).
\end{equation}
Samples with non-finite $D$ or $D\le 0$ are rejected. The TUR bound used in the plots is
\begin{equation}
\sigma_{\mathrm{TUR}}=\frac{j^2}{D}.
\end{equation}

\subsection{Full entropy production rate}
For diagnostic purposes, we also compute the full steady-state entropy production rate
\begin{equation}
\sigma=\sum_{i\neq j} p_j W_{ij}\,\ln\!\left(\frac{p_j W_{ij}}{p_i W_{ji}}\right),
\end{equation}
where terms with zero or non-finite arguments are skipped. This quantity is not used to determine point colors in the requested figures, but it is stored as additional output.

\subsection{Survival generator and conditional probabilities}
To compute the waiting-time based quantities for the single observed link, we introduce the survival generator $S$, obtained by removing the visible off-diagonal transition rates while leaving all diagonal entries unchanged,
\begin{equation}
S_{ij}=
\begin{cases}
0, & (i,j)=(1,4)\ \text{or}\ (4,1),\\
W_{ij}, & \text{otherwise}.
\end{cases}
\end{equation}
We compute $S^{-1}$ numerically, and evaluate the conditional probabilities of observing the same direction twice in a row,
\begin{equation}
P(+|+)= -\,W_{1,4}\,(S^{-1})_{4,1}, \qquad
P(-|-)= -\,W_{4,1}\,(S^{-1})_{1,4},
\end{equation}
where the index ordering follows the column convention. We reject samples for which either conditional probability is non-finite or nonpositive.

We also compute the traffic on the observed link and the probabilities of observing a $+$ or $-$ event,
\begin{equation}
h_K=p_4 W_{1,4}+p_1 W_{4,1}, \qquad
P(+)=\frac{p_4 W_{1,4}}{h_K}, \qquad
P(-)=\frac{p_1 W_{4,1}}{h_K},
\end{equation}
rejecting samples with non-finite or nonpositive $h_K$.

\subsection{Choice of time grid}
The waiting-time integrals are evaluated on a finite grid $t\in[t_{\min},t_{\max}]$. The upper cutoff $t_{\max}$ is chosen from the dominant decay rate of the survival dynamics. We compute the eigenvalues of $S$ and set
\begin{equation}
\alpha=-\max\{\mathrm{Re}\,\lambda: \lambda \text{ eigenvalue of } S\},
\qquad
t_{\max}=\frac{\texttt{tail\_factor}}{\alpha},
\end{equation}
requiring $\alpha>0$. A small positive $t_{\min}$ is used to avoid numerical issues at $t=0$. The grid is a uniform mesh,
\begin{equation}
t_k=t_{\min}+k\Delta t,\quad k=0,1,\dots,M-1,\qquad \Delta t=\frac{t_{\max}-t_{\min}}{M-1},
\end{equation}
with $M=\texttt{grid\_points}$.

We have used $\texttt{tail\_factor}=40.0, \, t_{\min}=10^{-12},\, \texttt{grid\_points}=8000$.
\subsection{Matrix exponential entries and waiting-time densities}
To evaluate the waiting-time densities, we need specific entries of $\exp(tS)$. We compute an eigendecomposition
\begin{equation}
S=V\Lambda V^{-1},
\end{equation}
numerically, where $\Lambda$ is diagonal with eigenvalues $\lambda_a$. Then
\begin{equation}
\exp(tS)=V\,\exp(t\Lambda)\,V^{-1},
\end{equation}
and the required matrix entry is obtained as
\begin{equation}
\big[\exp(tS)\big]_{ij}=\sum_{a} V_{i a}\,e^{\lambda_a t}\,(V^{-1})_{a j}.
\end{equation}
In floating-point arithmetic, small imaginary parts can arise, so we take the real part of the sum.

The waiting-time density for two successive $+$ events is evaluated on the grid as
\begin{equation}
\psi_{++}(t_k)= -\frac{\big[\exp(t_k S)\big]_{4,1}}{(S^{-1})_{4,1}},
\end{equation}
and similarly for two successive $-$ events,
\begin{equation}
\psi_{--}(t_k)= -\frac{\big[\exp(t_k S)\big]_{1,4}}{(S^{-1})_{1,4}}.
\end{equation}
Any negative values caused by numerical noise are clipped to zero. Each discrete density is then renormalized using the trapezoidal rule,
\begin{equation}
\psi(t_k)\leftarrow \frac{\psi(t_k)}{\sum_{k=0}^{M-2}\frac{\Delta t}{2}\big(\psi(t_k)+\psi(t_{k+1})\big)}.
\end{equation}
Samples for which normalization fails (non-finite or nonpositive integral) are rejected.

\subsection{Discrete KL divergence and $\sigma_t$}
The KL divergence between two waiting-time densities is evaluated by trapezoidal integration on the grid. For $D_{\mathrm{KL}}(\psi\Vert\phi)=\int dt\,\psi(t)\ln(\psi(t)/\phi(t))$, we use
\begin{equation}
D_{\mathrm{KL}}(\psi\Vert\phi)\approx \sum_{k=0}^{M-2}\frac{\Delta t}{2}\Big[g(t_k)+g(t_{k+1})\Big],
\qquad
g(t)=\psi(t)\ln\!\left(\frac{\psi(t)}{\phi(t)}\right).
\end{equation}
To avoid $\ln 0$ and division by zero, we replace each density by $\max(\text{density},\varepsilon)$ inside the logarithm, with a tiny $\varepsilon$ (machine-safe floor).
We have used $\varepsilon=10^{-300}$.

The waiting-time contribution is then computed as
\begin{equation}
\sigma_t
=h_K\Big[
P(+)P(+|+)\,D_{\mathrm{KL}}(\psi_{++}\Vert\psi_{--})
+
P(-)P(-|-)\,D_{\mathrm{KL}}(\psi_{--}\Vert\psi_{++})
\Big].
\end{equation}

\subsection{Embedded-chain term and $\sigma_{\mathcal L}$}
The embedded-chain contribution is computed as
\begin{equation}
\sigma_{\ell}=h_K\,[P(+)-P(-)]\,\ln\!\left(\frac{P(+|+)}{P(-|-)}\right),
\end{equation}
and the total waiting-time bound is
\begin{equation}
\sigma_{\mathcal L}=\sigma_{\ell}+\sigma_t.
\end{equation}
Samples are discarded if $\sigma_{\mathcal L}$ is non-finite or nonpositive.

\section{Double-check}
\label{app:doublecheck}

\subsection{Stationary distribution and inverse of survival generator}
We validate the computation of $ {\boldsymbol{p}}$ and ${S^{-1}}$ directly: 
\begin{equation}
\bm 1^{\mathsf T}{\bm p} = 1.0000000000000002,
\end{equation}
\begin{equation}
W\bm p \approx
\begin{pmatrix}
0\\
-1.78\times 10^{-15}\\
1.78\times 10^{-15}\\
0
\end{pmatrix},
\qquad
\|W\bm p\|_2 \approx 2.51\times 10^{-15},\qquad
\|W\bm p\|_\infty \approx 1.78\times 10^{-15},
\end{equation}
and
\begin{equation}
\big\|S{S^{-1}}-I\big\|_{\infty}\approx 2.22\times 10^{-16},
\qquad
\big\|{S^{-1}}S-I\big\|_{\infty}\approx 2.57\times 10^{-16}.
\end{equation}

\subsection{Diffusion constant via finite differences of the tilted dominant eigenvalue}
To validate the numerical precision of $D$, we performed an independent calculation based on finite differences of $\lambda(k)$, the dominant eigenvalue of the tilted generator. The crucial point is that this check must use the same tilting convention as the perturbation method: only the two observed off-diagonal rates are multiplied by exponential factors, while all diagonal elements are kept unchanged. Specifically, for the observed link,
\begin{equation}
W(k)_{1,4}=W_{1,4}e^{k},\qquad
W(k)_{4,1}=W_{4,1}e^{-k},
\end{equation}
and all other matrix elements, including all diagonal entries, are identical to those of $W$.

For each step size $h$, we computed $\lambda(h)$ and $\lambda(-h)$ and formed central finite differences,
\begin{equation}
j_{\mathrm{fd}}(h)=\frac{\lambda(h)-\lambda(-h)}{2h},
\qquad
(2D)_{\mathrm{fd}}(h)=\frac{\lambda(h)-2\lambda(0)+\lambda(-h)}{h^2},
\qquad
D_{\mathrm{fd}}(h)=\frac{1}{2}(2D)_{\mathrm{fd}}(h).
\end{equation}
The numerical results were:
\begin{equation}
\begin{array}{c c c c}
\hline
h & j_{\mathrm{fd}}(h) & D_{\mathrm{fd}}(h) & \text{relative error in } D\\
\hline
10^{-2} & -5.341000826513 & 2.309315113784 & 1.03\times 10^{-6}\\
10^{-3} & -5.340998703055 & 2.309312755244 & 6.83\times 10^{-9}\\
10^{-4} & -5.340998681831 & 2.309306523784 & -2.69\times 10^{-6}\\
10^{-5} & -5.340998682613 & 2.309477054041 & 7.12\times 10^{-5} \\ 
\hline
\end{array}
\end{equation}
The $h=10^{-3}$ case provides a clear ``plateau'' where both $j_{\mathrm{fd}}$ and $D_{\mathrm{fd}}$ agree with the perturbation results at the level of $\sim 10^{-9}$ relative error for $D$ (and even smaller for $j$). For $h=10^{-2}$, the larger discrepancy is consistent with truncation error in the second-derivative approximation. For $h=10^{-5}$, the observed drift (relative error $\sim 10^{-4}$) is consistent with round-off and eigenvalue-solver sensitivity when subtracting nearly equal quantities at very small step sizes.

The visible traffic and the probabilities of observing $\pm$ are
\begin{equation}
h_K=p_4 k_+ + p_1 k_- = 15.9578515161254,
\qquad
P(+)=\frac{p_4 k_+}{h_K}=0.332652952177398,
\qquad
P(-)=\frac{p_1 k_-}{h_K}=0.667347047822602.
\end{equation}
The conditional probabilities of observing the same direction twice in a row are computed from $S^{-1}$ as
\begin{equation}
P(+|+)=-k_+(S^{-1})_{4,1}=0.0565635811515161,
\qquad
P(-|-)= -k_-(S^{-1})_{1,4}=0.529724584913357,
\end{equation}
which are positive and less than $1$, as required.

The embedded-chain contribution is
\begin{equation}
\sigma_\ell
= h_K\,[P(+)-P(-)]\ln\!\left(\frac{P(+|+)}{P(-|-)}\right)
= 11.9477707076164.
\end{equation}

\subsection{Double check of $\sigma_t$ and $\sigma_{\mathcal L}$ for the counterexample}

 The waiting-time contribution $\sigma_t$ is computed from the waiting-time densities $\psi_{++}(t)$ and $\psi_{--}(t)$.
Using the standard identity valid when $\mathrm{Re}\,\lambda(S)<0$,
\begin{equation}
\psi_{++}(t)= -\frac{[\exp(tS)]_{4,1}}{(S^{-1})_{4,1}},
\qquad
\psi_{--}(t)= -\frac{[\exp(tS)]_{1,4}}{(S^{-1})_{1,4}},
\end{equation}
we evaluate these functions on a uniform time grid $t\in[t_{\min},t_{\max}]$ with $t_{\min}=10^{-12}$ and
\begin{equation}
t_{\max}=\frac{\texttt{tail\_factor}}{\alpha},
\qquad
\alpha=-\max\mathrm{Re}\,\lambda(S)=14.9936792074212.
\end{equation}
With \texttt{tail\_factor}$=40$, this gives $t_{\max}=2.66779083683488$.
On this grid we clip small negative values due to round-off, and then normalize by trapezoidal integration.
For the high-resolution run reported below, the unnormalized integrals over $[t_{\min},t_{\max}]$ satisfy
\begin{equation}
\int_{t_{\min}}^{t_{\max}}\psi_{++}(t)\,dt = 0.999999999999996,
\qquad
\int_{t_{\min}}^{t_{\max}}\psi_{--}(t)\,dt = 0.999999999999998,
\end{equation}
showing that the finite cutoff already captures essentially all probability mass.

We performed two independent checks.
First, we tested convergence with respect to the time discretization and the choice of $(t_{\min},t_{\max})$.
Second, we computed the required entries of $\exp(tS)$ in two different ways,
by an eigendecomposition entry formula and by direct evaluation using a matrix exponential routine, and confirmed agreement.
Representative results are shown below, where ``eig'' denotes the eigendecomposition-based evaluation of $[\exp(tS)]_{ij}$ and ``expm'' denotes direct computation of $\exp(tS)$.

\begin{equation}
\begin{array}{c c c cc}
\hline
\text{method} & \texttt{tail\_factor} & \texttt{grid\_points} & \sigma_t & \sigma_{\mathcal L}\\
\hline
\text{eig}  & 40 & 8000   & 0.256306400075445 & 12.2040771076919\\
\text{eig}  & 40 & 20000  & 0.256306401477457 & 12.2040771090939\\
\text{eig}  & 40 & 80000  & 0.256306401514138 & 12.2040771091306\\
\text{expm} & 40 & 20000  & 0.256306401477453 & 12.2040771090939\\
\text{eig}  & 60 & 8000   & 0.256306394231033 & 12.2040771018475\\
\text{eig}  & 60 & 200000 & 0.256306401514261 & 12.2040771091307 \\
\hline 
\end{array}
\end{equation}

These checks show the following.
For \texttt{tail\_factor}$=40$, already \texttt{grid\_points}$=8000$ yields $\sigma_t$ within about $1.4\times 10^{-9}$ of the high-resolution value, and increasing the grid rapidly stabilizes $\sigma_t$ and hence $\sigma_{\mathcal L}$.
When the time interval is enlarged (\texttt{tail\_factor}$=60$) without increasing grid resolution, the coarser discretization introduces a visible error, but the value converges back to the same limit once the grid is refined.
At fixed parameters, the ``eig'' and ``expm'' evaluations agree to within numerical round-off, indicating that $\sigma_t$ is not sensitive to the particular numerical method used to compute $\exp(tS)$.

For the counterexample matrix $W$, the waiting-time term and the bound are numerically well validated:
\begin{equation}
\sigma_t = 0.256306401514275,
\qquad
\sigma_{\mathcal L} = 12.2040771091307,
\end{equation}
with the remaining numerical uncertainty dominated by time discretization and shown by the convergence tests above to be at most on the order of $10^{-9}$ for $\sigma_t$.

\bibliography{ref}

\end{document}